\documentclass[12pt]{article}
\usepackage{epsfig}
\usepackage{color}
\usepackage{amssymb,amsmath}
\usepackage{graphicx}
\usepackage{epsfig}

\newcommand{\bea}{\begin{eqnarray}}
\newcommand{\eea}{\end{eqnarray}}

 \makeatletter
\def\alt{\mathrel{\mathpalette\gl@align<}}
\def\agt{\mathrel{\mathpalette\gl@align>}}
\def\gl@align#1#2{\lower.6ex\vbox{\baselineskip\z@skip\lineskip\z@
\ialign{$\m@th#1\hfil##\hfil$\crcr#2\crcr\sim\crcr}}} \makeatother

\begin{document}
\begin{flushright}
KEK-TH-1578 \\
UT-12-32
\end{flushright}
\vspace*{1.0cm}

\begin{center}
\baselineskip 20pt 
{\Large\bf 
Null Radiation Zone at the LHC
}
\vspace{1cm}

{\large 
Kaoru Hagiwara$^{a}$, Toshifumi Yamada$^{b}$
} \vspace{.5cm}

{\baselineskip 20pt \it

$^{a}$ KEK Theory Center and SOKENDAI, \\
1-1 Oho, Tsukuba, Ibaraki 305-0801, Japan \\ \
\\

$^{b}$ Department of Physics, University of Tokyo, \\
7-3-1 Hongo, Bunkyo-ku, Tokyo 113-0033, Japan }

\vspace{.5cm}

\vspace{1.5cm} {\bf Abstract} \end{center}

The null radiation zone theorem states that,
 when special kinematical conditions are satisfied,
 all the helicity amplitudes of a parton-level subprocess where a vector current is emitted
 vanish due to destructive interference among different diagrams.
We study the manifestation of the theorem in $pp$ collisions at the $\sqrt{s}=8$ TeV LHC.
The theorem predicts that the cross section for $p p \rightarrow j j \gamma$ events
 is suppressed when the transverse momenta of the two jets are similar
 and when the rapidity difference between the photon and the cluster of the jets is nearly zero,
 because the $u u \rightarrow u u \gamma$ subprocess,
 which dominates in events with large $j j \gamma$ invaraint mass,
 has strong destructive interference
 in this region.
We confirm this prediction by the calculation with MadGraph 5,
 and show that the suppression on the $p p \rightarrow j j \gamma$ cross section
 is observable at the LHC.

\thispagestyle{empty}

\newpage

\setcounter{footnote}{0}
\baselineskip 18pt
%

\ \ \ One of the goals of the LHC is to
 confirm predictions of the standard model with its unprecedented energy and luminosity.
In this letter, we discuss the null radiation zone theorem \cite{null zone theorem, null zone} and
 its manifestation in hard events at the LHC.
We especially focus on the following parton-level subprocesses:
\begin{eqnarray}
q \ q^{\prime} &\rightarrow& q \ q ^{\prime} \ \gamma \ , \ \ \ \ \ \ \ ( \ q, q^{\prime} = u, d \ ) 
\label{qqg}
\end{eqnarray}
 which contribute to events with two hard jets and a photon.
The four contributing diagrams are shown in Figure 1.
The null radiation zone theorem relates the charges of the quarks with
 the kinematic conditions under which the diagrams of a subprocess have 
 strong destructive or constructive interference.
More generally,
 the theorem holds at the classical level when the photon
 minimally couples to the particles and its anomalous-magnetic-moment
 g-2, quadrapole moments, etc. are absent.
Early discussions on its implications in hadron collisions
 are found in refs.~\cite{had}, and
 those at HERA $e p$ collider experiments
 are found in refs.~\cite{hera}.

The null radiation zone theorem in its general form \cite{null zone theorem, ivdlhc} 
 is described as follows.
Consider the following process in which an Abelian vector current, $V$, is emitted:
\begin{eqnarray}
a \ + \ b &\rightarrow& 1 \ + \ 2 \ + \ V \ .
\end{eqnarray}
The invariant mass for the vector current $V$ is either zero or non-zero.
We label the particle four-momenta and charges in the initial state
 by $p_i, \ Q_i$ $(i=a,b)$,
 those in the final state by $p_f, \ Q_f$ $(f=1,2)$,
 and the four-momentum of the vector current by $p_V$.
Then the theorem states that the tree-level scattering amplitude should vanish for all helicities
 when the following conditions are satisfied:
\begin{eqnarray}
\frac{Q_i}{2 p_i \cdot p_V - p_V^2} &=& \frac{Q_f}{2 p_f \cdot p_V + p_V^2} \ = \ {\rm (a \ common \ value)}
\ \ \ {\rm for \ all} \ i \ {\rm and} \ f \ ,  \label{cond}
\\
\sum_i Q_i &=& \sum_f Q_f \ . \label{cond2}
\end{eqnarray}
We note here that the latter condition (\ref{cond2}) dictates the charge conservation,
 while the conditions (\ref{cond}) have a solution in the physical region
 only when all $Q_i$ and $Q_f$ have the same sign and when the vector boson is massless, $p_V^2=0$.
The generalized form is useful to identify kinematical regions where the amplitudes with a vetor boson emission
 interfere destructively or constructively \cite{ivdlhc}.

\begin{figure}[tbp]
  \begin{center}
   \includegraphics[width=140mm]{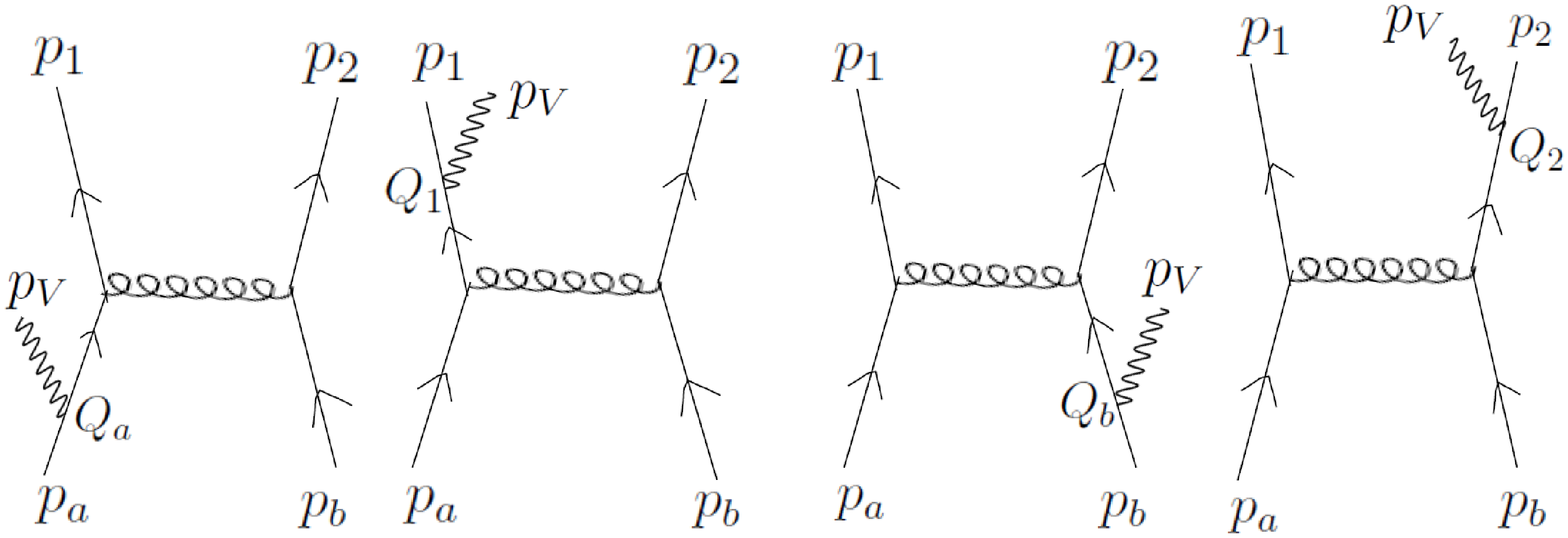}
  \end{center}
 \caption{Feynman diagrams for a subprocess
  $q q^{\prime} \rightarrow q q^{\prime} \gamma$
  $(q,q^{\prime}=u,d)$.
 When $q=q^{\prime}$, anti-symmetrization of the final state quarks should be performed.
 }
\end{figure}

Let us focus on the Feynman diagrams depicted in Figure 1,
 where the sum of the left two diagrams and that of the right two
 are respectively gauge invariant.
We assume that the quarks are all massless: 
$p_a^2 = p_b^2 = p_1^2 = p_2^2 = 0$.
If the charges of the quarks are the same, i.e.,
 $Q_a = Q_b = Q_1 = Q_2$ holds,
 the condition for the null radiation zone eq.~(\ref{cond})
 reduces to the following simple relations:
\begin{eqnarray}
y(p_1+p_2) \ - \ y(p_V) &=& 0 \ , \label{rap}
\\
p_{T1} &=& p_{T2} \ , \label{pt}
\\
p_V^2 &=& 0 \ , \label{q2}
\end{eqnarray} 
 where $y(p_1+p_2)$ denotes the rapidity of the four-momentum sum of
 the final state quarks 1 and 2,
 and $y(p_V)$ denotes the rapidity of the vector current $V$.
$p_{T1}$ and $p_{T2}$ denote the transverse momenta of the final state quarks 1 and 2.
 (The three-momenta of $p_a$ and $p_b$ are assumed to be along the z-axis.)
In other words, when the conditions eqs.~(\ref{rap}, \ref{pt}, \ref{q2}) are satisfied,
 the diagrams of Figure 1 have maximal destructive interference
 and their amplitudes sum up to zero.
Consequently, if $Q_a=Q_1$ and $Q_b=Q_2$ hold,
 but $Q_a$ and $Q_b$ are not necessarily equal,
 the sum of the amplitudes is proportional to $(Q_a-Q_b)$
 and the cross section to $(Q_a-Q_b)^2$
 in the kinematic region satisfying eqs.~(\ref{rap}, \ref{pt}, \ref{q2}).
Were it not for interference terms between the left two and the right two diagrams of Figure 1,
 the cross section would be given by $Q_a^2+Q_b^2$.
We thus notice that, in the region satisfying eqs.~(\ref{rap}, \ref{pt}, \ref{q2}),
 destructive interference occurs when $Q_a$ and $Q_b$ take the same sign
 and constructive interference occurs when they take the opposite signs.
\footnote{
We can directly confirm the appearance of destructive and constructive interferences
 from the explicit formulae found in eqs.~(11, 12) of ref.~\cite{brems}
 with the replacement of $C_1, \ C_2, \ C_3$ and $C_4$ with appropriate electric charges.
}
\\

We check the prediction of the null radiation zone theorem by simulating parton-level processes
 with MadGraph 5 \cite{mg}.
We calculate the cross sections for
 $u u \rightarrow u u \gamma$ subprocess in $pp$ collisions at $\sqrt{s}=8$ TeV
 with and without the interference terms between the left two and the right two diagrams of Figure 1.
For comparison, we also calculate the cross sections for 
 $u \bar{u} \rightarrow u \bar{u} \gamma$ subprocess in $p\bar{p}$ collisions at $\sqrt{s}=8$ TeV.
The up quark and anti-up quark have opposite electric charges, and 
 the parton distribution function of the up quark in a proton
 is the same as that of the anti-up quark in an anti-proton.
Although $s$-channel exchange of a gluon contributes 
 to $u \bar{u} \rightarrow u \bar{u} \gamma$ subprocess
 in addition to $t$-channel exchange depicted in Figure 1,
 such contribution can be suppressed by the selection cut which we introduce later.
For these subprocesses, the condition eq.~(\ref{q2}) is automatically fulfilled.
The theorem predicts that,
 when the conditions eq.~(\ref{rap}, \ref{pt}) are nearly satisfied,
 the cross section for $u u \rightarrow u u \gamma$ subprocess in $pp$ collisions
 is strongly suppressed compared to the cross section estimated without the interference terms.
On the other hand,
 the cross section for $u \bar{u} \rightarrow u \bar{u} \gamma$ subprocess in $p\bar{p}$ collisions
 is expected to be enhanced compared to the estimate without the interferene terms,
 because the up and anti-up quarks have opposite electric charges
 and hence the contributing diagrams have constructive interference.

\begin{figure}[htbp]
 \begin{minipage}{0.5\hsize}
  \begin{center}
   \includegraphics[width=85mm]{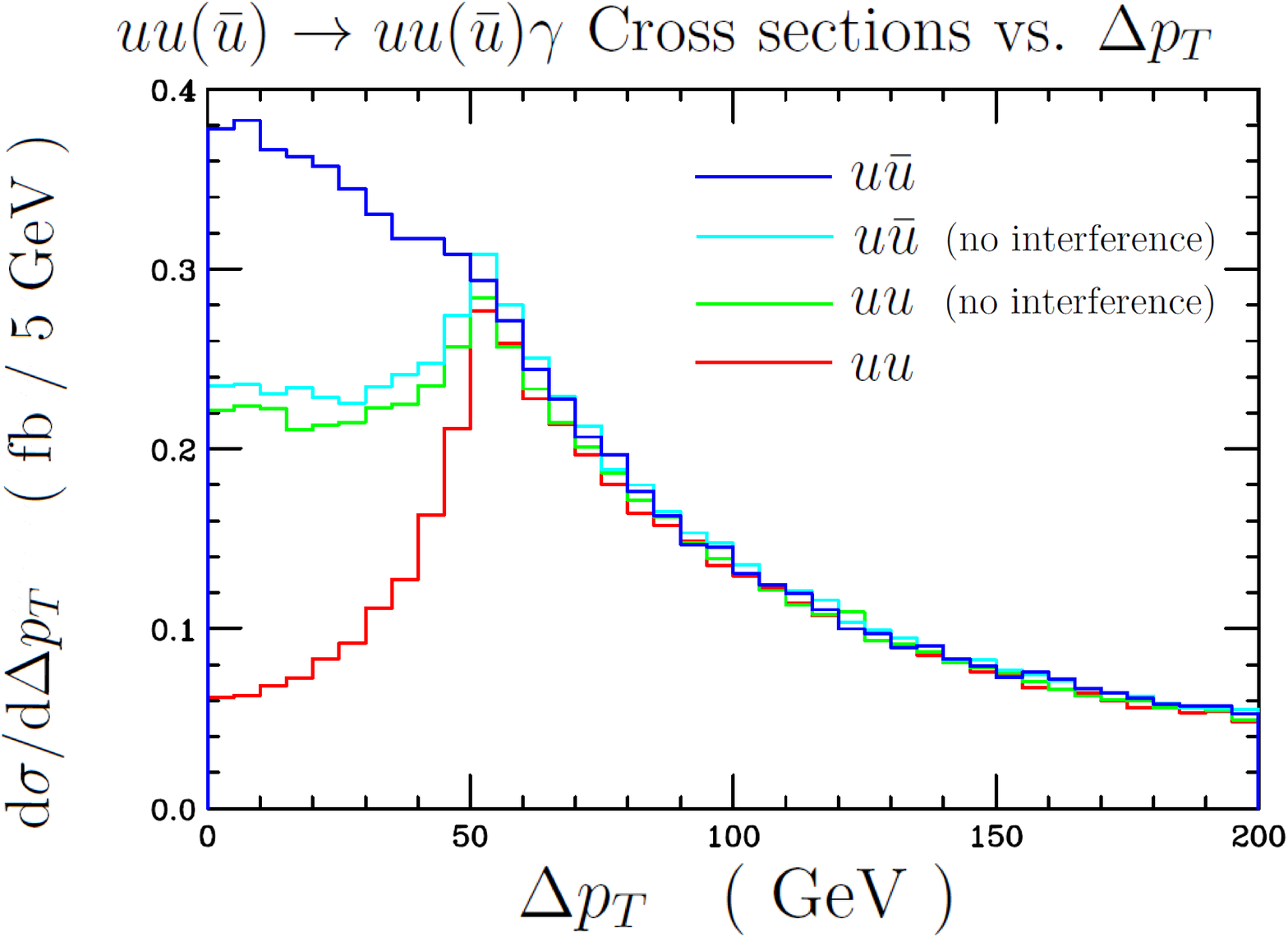}
  \end{center}
 \end{minipage}
 \begin{minipage}{0.5\hsize}
  \begin{center}
   \includegraphics[width=80mm]{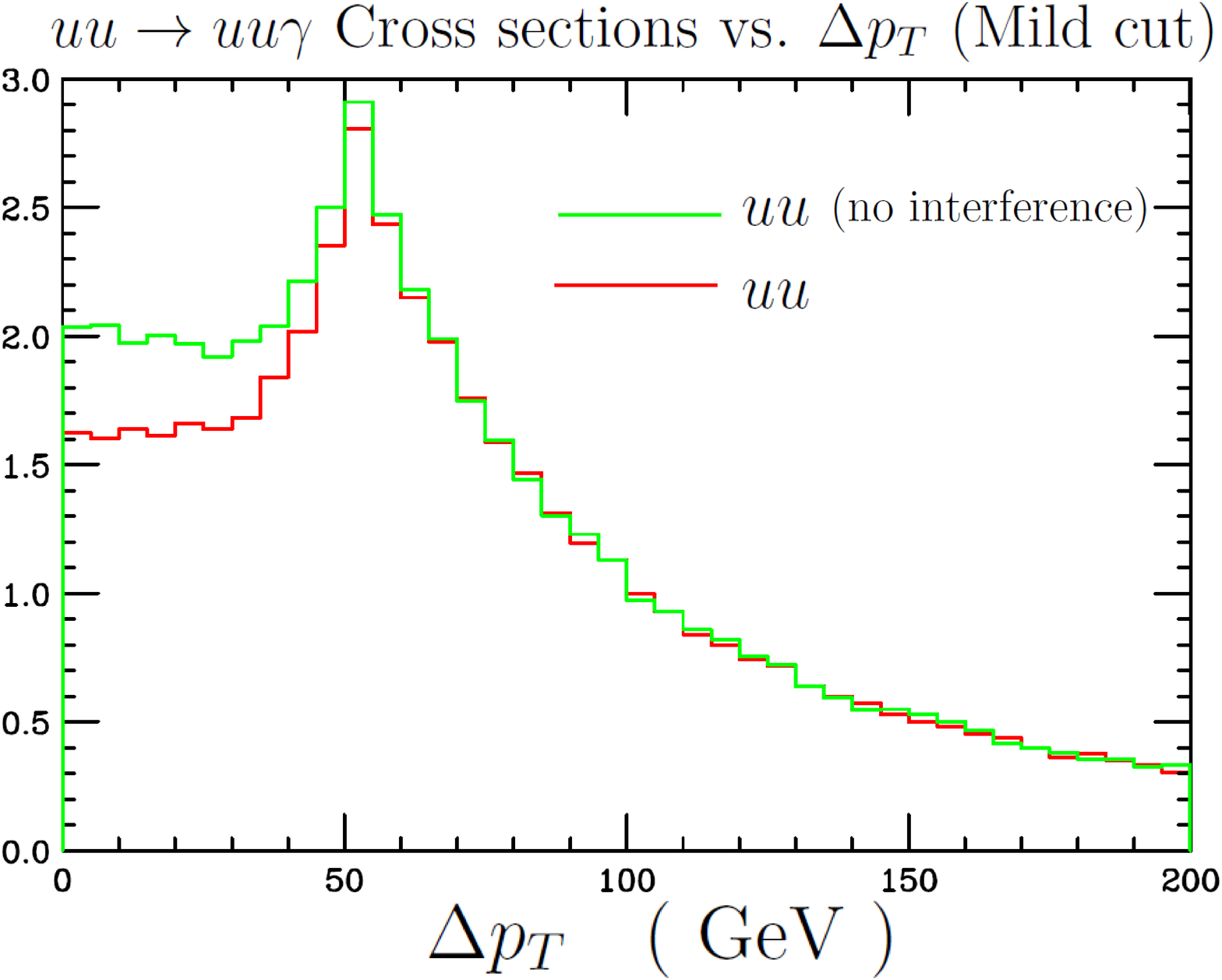}
  \end{center}
 \end{minipage}
 \caption{(Left) The differential cross sections, d$\sigma/$d$\Delta p_T$ ($\Delta p_T = \vert p_{T1}-p_{T2} \vert$),
  for the $u u \rightarrow u u \gamma$ subprocess 
  in $pp$ collisions at $\sqrt{s}=8$ TeV with the cut eqs.~(a, b, c, d).
 The interference terms between the left two and the right two diagrams of Figure 1
  are included for the lower red line, and are dropped for the lower-middle green line.
 Also shown are the differential cross sections for
  the $u \bar{u} \rightarrow u \bar{u} \gamma$ subprocess in $p\bar{p}$ collisions at $\sqrt{s}=8$ TeV
  with the interference terms included (upper, blue), and dropped (upper-middle, cyan).
 (Right) The same as the left graph for the $u u \rightarrow u u \gamma$ contribution,
  but the requirement eq.~(d) is mildened to $\Delta y < 2.0$ (d$^{\prime}$).
 }
\end{figure}

Figure 2 shows the cross sections for 
 the $u u \rightarrow u u \gamma$ subprocess in $pp$ collisions at $\sqrt{s}=8$ TeV
 where the interference terms between the left two and the right two diagrams of Figure 1
 are included or dropped.
Also shown are the cross sections for
 the $u \bar{u} \rightarrow u \bar{u} \gamma$ subprocess in $p\bar{p}$ collisions at $\sqrt{s}=8$ TeV.
Each cross section is plotted as a function of the difference between the transverse momenta of the two jets,
 $\Delta p_T \equiv \vert p_{T1} - p_{T2} \vert$.
For the left graph of Figure 2, the kinematical cut is as follows:
\begin{itemize}
 \item Require two jets with $\vert \eta \vert < 4.5$ and $p_{T} > 25$~GeV. \hspace{\fill} (a)
 \item Require a photon with $\vert \eta_{\gamma} \vert < 1.44$ or $1.57 < \vert \eta_{\gamma} \vert < 2.5$ 
  and $p_{T \gamma} > 50$~GeV. \hspace{\fill} (b)
 \item Require $M_{jj\gamma} \equiv \sqrt{ (p_{j1}+p_{j2}+p_{\gamma})^2 } > 2.5$ TeV,
  where $p_{j1}, \ p_{j2}$ and $p_{\gamma}$ denote the four-momenta of the jets ($p_{Tj1} > p_{Tj2}$)
  and the photon, respectively. \hspace{\fill} (c)
 \item Require $\Delta y \equiv \vert y(p_{j1}+p_{j2}) - \eta_{\gamma} \vert < 0.75$,
  where $y(p_{j1}+p_{j2})$ denotes the rapidity of the four-momentum sum of the two leading jets.
  \hspace{\fill} (d)
\end{itemize}
The requirement (d) corresponds to the condition eq.~(\ref{rap}) for the null radiation zone.
For the right graph of Figure 2, we have mildened the requirement eq.~(d) to
\begin{eqnarray*}
\Delta y &=& \vert y(p_{j1}+p_{j2}) - \eta_{\gamma} \vert \ < \ 2.0 \ . \ \ \ \ \ \ \ \ \ \ \ \ \ \ \ \ \ \ \ ({\rm d}^{\prime})
\end{eqnarray*}

We find in the left graph that, in the region with small $\Delta p_T$,
 the $u u \rightarrow u u \gamma$ cross section with the interference terms included 
 is significantly suppressed compared to that with the interference terms dropped,
 while the opposite is the case for the $u \bar{u} \rightarrow u \bar{u} \gamma$ cross section.
By comparing the right graph with the left one,
 we notice that the severe cut on $\Delta y$, $\Delta y < 0.75$ in eq.~(d),
 is effective for confirming the null radiation zone in the small $\Delta p_T$ region.
We thus confirm that, around the region 
 with $y(p_{j1}+p_{j2}) - \eta_{\gamma} \simeq 0$ and $p_{T1} \simeq p_{T2}$, i.e., 
 when the conditions eq.~(\ref{rap}, \ref{pt}) are nearly satisfied,
 the $u u \rightarrow u u \gamma$ subprocess has strong destructive interference,
 whereas the $u \bar{u} \rightarrow u \bar{u} \gamma$ process has strong constructive interference.
\\

We next calculate the cross sections for the $p p \rightarrow j j \gamma$ process at $\sqrt{s}=8$ TeV
 with and without the interference terms between the left two and the right two diagrams of Figure 1
 by using MadGraph 5 \cite{mg}.
We also calculate the cross sections for the $p \bar{p} \rightarrow j j \gamma$ process at $\sqrt{s}=8$ TeV
 as a reference.

In Table 1, we summarize the cross sections for parton-level subprocesses of the type
 $q q^{\prime} (\bar{q}^{\prime}) \rightarrow q q^{\prime} (\bar{q}^{\prime}) \gamma$
 in the null radiation zone 
 inside the kinematical region defined by eqs.~(\ref{rap}, \ref{pt}, \ref{q2})
 based on the $(Q_a-Q_b)^2$ counting rule, where
 $Q_a$ and $Q_b$ are the electric charges of the colliding partons.
Shown in the right column are the corresponding values of $Q_a^2+Q_b^2$,
 which would be the expected cross sections in the absence of 
 the interference terms.
\begin{table}
\begin{center}
\begin{tabular}{|c|c|c|} \hline
Subprocess      &  Cross section with Int. & Cross section without Int. \\ \hline
$ u u $         &  0                       & $8/9 \ C$              \\ \hline
$ d d $         &  0                       & $1/9 \ C$                \\ \hline
$ u d , \ d u $    &  $C$                     & $5/9 \ C$                 \\ \hline
$ u \bar{u} $    &  $16/9 \ C$              & $8/9 \ C$                    \\ \hline
$ d \bar{d} $    &  $4/9 \ C$               & $1/9 \ C$             \\ \hline
$ u \bar{d}, \ d \bar{u} $   &  $1/9 \ C$               & $5/9 \ C$        \\ \hline
\end{tabular}
\end{center}
\caption{Estimates on the cross sections for quark-quark and quark-anti-quark subprocesses
 in the kinematical region satisfying eqs.~(\ref{rap}, \ref{pt}, \ref{q2}),
 with and without the interference terms.
$C$ is a common number.
Contribution of each subprocess to the $j j \gamma$ production in $pp$ and $p\bar{p}$ collisions
 will be obtained by multiplying the relevant parton distribution functions.
Contributions of $s$-channel gluon exchange to $u \bar{u}$ and $d \bar{d}$ collisions are neglected.
}
\end{table}

By using the counting rule of Table 1, 
 we estimate the effect of the interference in the $pp \rightarrow j j \gamma$ process as follows.
Let us assume for simplicity that contributions from sea quarks are negligibly small.
Then $pp$ collisions have contributins from $uu$, $dd$ and $ud+du$ collisions,
 whereas $p\bar{p}$ collisions are made of $u\bar{u}$, $d\bar{d}$ and $u\bar{d}+d\bar{u}$ collisions.
These contributions are proportional to the parton-parton luminosity functions
\begin{eqnarray}
L_{ab}(Q) &=& \int {\rm d}x_a \ \int {\rm d}x_b \ D_{a/p}(x_a, Q) \ D_{b/p}(x_b, Q) \
\theta \left( x_a x_b - \left( \frac{2.5 \ {\rm TeV}}{8 \ {\rm TeV}} \right)^2 \right) \ , \label{plf}
\end{eqnarray}
 where $D_{a/p}$ denotes the parton distribution function (PDF) for parton $a$ in a proton
 and $Q$ denotes the renormalization scale.
From eq.~(\ref{plf}), we find 
\begin{eqnarray}
L_{dd} / L_{uu} &=& 0.11 \ , \\
(L_{ud}+L_{du}) / L_{uu} &=& 0.68
\end{eqnarray}
 for PDF CTEQ6L1 \cite{cteq6l1} and $Q=p_{T}^{\rm{cut}}=25$ GeV.
In the approximation of neglecting $s$-channel gluon exchange amplitude,
 we can use the same luminosity function ratios for $u \bar{u}, \ d \bar{d}$ and $u \bar{d} + d \bar{u}$
 contributions in $p \bar{p}$ collisions.
The cross section for the $p p \rightarrow j j \gamma$ process with the interference terms included
 is then given by
\begin{eqnarray}
L_{uu} \cdot 0 \ + \ (L_{ud}+L_{du}) \cdot C \ + \ L_{dd} \cdot 0 &=& 0.68 \ L_{uu} C
\end{eqnarray}
 where the first term corresponds to the contribution of $uu$ subprocess, 
 the second term to $ud$ subprocess and the third term to $dd$ subprocess.
Note that the null radiation zone theorem holds for both $uu$ and $dd$ collision processes
 where all the charges in eq.~(\ref{cond}) are the same,
 whereas for $ud$ collision process, the interference is constructive because
 $(Q_u - Q_d)^2 - (Q_u^2 + Q_d^2) = -2 Q_u Q_d > 0$.
On the other hand, when the interference terms were dropped,
 the cross section would be given by
\begin{eqnarray}
L_{uu} \cdot (8/9) C \ + \ (L_{ud}+L_{du}) \cdot (5/9) C \ + \ L_{dd} \cdot (1/9) C &=& 1.3 \ L_{uu} C \ , \label{ppNI}
\end{eqnarray}
Since $0.68 L_{uu} C$ is about a half of $1.3 L_{uu} C$, 
 there is a hope that the destructive interference in the $p p \rightarrow j j \gamma$ process
 can be confirmed in the region satisfying eqs.~(\ref{rap}, \ref{pt}, \ref{q2}).

For the $p \bar{p} \rightarrow j j \gamma$ process,
 the cross section with the interference terms included
 is given by
\begin{eqnarray}
L_{uu} \cdot (16/9) C \ + \ (L_{ud}+L_{du}) \cdot (1/9) C \ 
+ \ L_{dd} \cdot (4/9) C &=& 1.9 \ L_{uu} C \ ,
\end{eqnarray}
 whereas the cross section without the interference terms
 would be given by
\begin{eqnarray}
4 \cdot (8/9) C \ + \ 4 \cdot (5/9) C \ 
+ \ 1 \cdot (1/9) C &=& 1.3\ L_{uu} C \ . \label{ppbarNI}
\end{eqnarray}
If it were not for the interference effects,
 the cross section for the $p \bar{p} \rightarrow j j \gamma$ process, eq.~(\ref{ppbarNI}),
 would be identical to that for the $p p \rightarrow j j \gamma$ process, eq.~(\ref{ppNI}).
The interference is now constructive due to the dominance of the $u \bar{u} \rightarrow u \bar{u} \gamma$ subprocess,
 and the difference between the $p \bar{p} \rightarrow j j \gamma$ and  $p p \rightarrow j j \gamma$ cross sections
 can be as large as a factor of 3.

Under the cut of only eqs.~(a, b, d),
 quark-gluon subprocesses ($q g \rightarrow q g \gamma$) dominate
 over $q q^{\prime} \rightarrow q q^{\prime} \gamma$ subprocesses
 for the $p p \rightarrow j j \gamma$ cross section.
To obtain the prediction of the null radiatiom zone theorem in the presence of the gluon background,
 we implement a severe cut on the invariant mass of the final-state particles
 so that we can select quark-quark subprocesses in $pp$ collisions,
 which have relatively large center-of-mass energy,
 while reducing the contributions from quark-gluon subprocesses.
So we use the cut of eqs.~(a, b, c, d), where eq.~(c) is to reduce the gluon background.
Figure 3 shows the cross sections for the $p p \rightarrow j j \gamma$ process
 where the interference terms between the left two and the right two diagrams of Figure 1
 are included or dropped.
Each cross section is plotted as a function of the difference between the transverse momenta of the two jets
 $\Delta p_T = \vert p_{T1} - p_{T2} \vert$.
In the top part of the figure,
 the cut eq.~(d) is replaced by a milder version eq.~(d$^{\prime}$) (replacing 0.75 by 2.0).

\begin{figure}[htbp]
  \begin{center}
   \includegraphics[width=135mm]{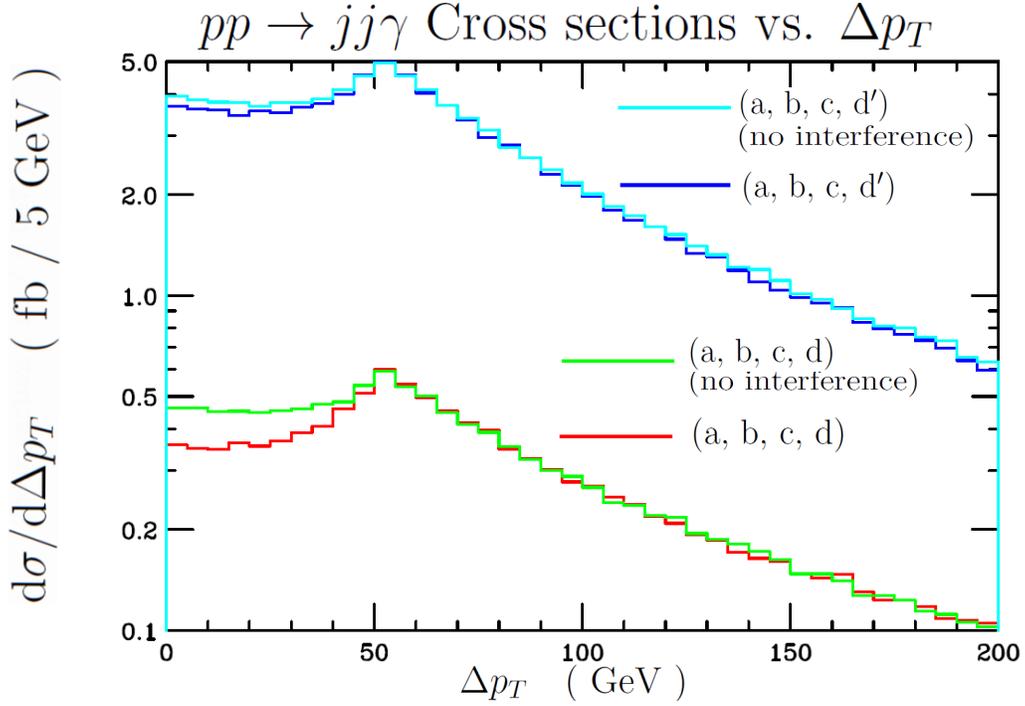}
  \end{center}
 \caption{The differential cross sections, d$\sigma$/$\Delta p_T$ ($\Delta p_T = \vert p_{T1}-p_{T2} \vert$),
  for the $p p \rightarrow j j \gamma$ process
  at $\sqrt{s}=8$ TeV with the cut eqs.~(a, b, c, d) for the red (lower) and green (lower-middle) lines,
  and with the cut eqs.~(a, b, c, d$^{\prime}$) for the blue (upper-middle) and cyan (upper) lines.
 The interference terms between the left two and the right two diagrams of Figure 1
  are included for the red and blue lines, and are dropped for the green and cyan lines.
 }
\end{figure}

We find from Figure 3 that, in the region with small $\Delta p_T$,
 the $p p \rightarrow j j \gamma$ cross section with the interference terms included 
 is suppressed compared to that with the interference terms dropped.
By comparing the upper two lines with the lower ones,
 we notice that a severe cut on $\Delta y$ helps extracting the effect of the interference.
We thus confirm that,
 around the kinematical region with $y(p_{j1}+p_{j2}) - \eta_{\gamma} \simeq 0$ and $p_{T1} \simeq p_{T2}$,
 the cross section for the $p p \rightarrow j j \gamma$ process is suppressed
 because of destructive interference between the left two and the right two diagrams of Figure 1.
This is in accord with our expectation based on the null radiation zone theorem.
\\

To conclude,
 it may be possible to observe the effect of the destructive interference in the $p p \rightarrow j j \gamma$ process
 at the $\sqrt{s}=8$ TeV LHC.
\footnote{
 We have examined the possibility of observing the constructive interference
 in the $p \bar{p} \rightarrow j j \gamma$ process at $\sqrt{s}=1.96$ TeV,
 but found it rather difficult mainly because very low $p_T$ cut ($\sim 6$ GeV)
 is required to gain the rate.}
Although the leading order matrix element level predictions of MadGraph may not give the right
 normalization of the cross sections,
 we expect that suppression of events with small $\Delta p_T$ 
 under the cut eqs.~(a, b, c, d) 
 can be observed through cross section ratios.
Our simulation predicts that the $p p \rightarrow j j \gamma$ cross section
 with the cut eqs.~(a, b, c, d) with $\Delta p_T < 40$ GeV
 and that with $\Delta p_T > 40$ GeV have the ratio of
\begin{eqnarray}
\frac{ \sigma_{pp}({\rm cut \ (a,b,c,d)}, \ \Delta p_T < 40 \ {\rm GeV}) }
{ \sigma_{pp}({\rm cut \ (a,b,c,d)}, \ \Delta p_T > 40 \ {\rm GeV}) }
 &=& \frac{ 2.936 \ {\rm fb} }{ 12.507 \ {\rm fb} } \ = \ 0.235 \ .
\end{eqnarray}
On the other hand, if the interference between the left two and the right two diagrams of Figure 2
 is dropped,
 the same ratio becomes
\begin{eqnarray}
\frac{ \sigma_{pp}^{\rm NI}({\rm cut \ (a,b,c,d)}, \ \Delta p_T < 40 \ {\rm GeV}) }
{ \sigma_{pp}^{\rm NI}({\rm cut \ (a,b,c,d)}, \ \Delta p_T > 40 \ {\rm GeV}) }
 &=& \frac{ 3.660 \ {\rm fb} }{ 12.664 \ {\rm fb} } \ = \ 0.289 \ .
\end{eqnarray}
Taking only statistical error into account,
 we can observe the contributions of the interference terms to $p p \rightarrow j j \gamma$ cross section
 at 2 $\sigma$ level
 with the integrated luminosity of 20 fb$^{-1}$.
If we adopt the milder cut of $\Delta y < 2.0$ (d$^{\prime}$) instead of eq.~(d),
 the above ratios respectively become 0.350 and 0.362,
 giving only about 1.3 $\sigma$ difference for the integrated luminosity of 20 fb$^{-1}$.
As expected, the severe cut of $\Delta y < 0.75$ (d) is useful to identify the strong destructive interference
 around the null radiation zone.

If the destructive interference is not observed, 
 it indicates that either there exists an anomalous coupling of a photon to quarks
 that breaks the premise of the null radiation zone theorem, or,
 more likely, our understanding of hadronic jets is inadequate.

Realistic event simulation with parton showering and hadronization 
 as well as QCD NLO corrections to the shape and the rate of the $\Delta p_T$ distributions
 are beyond the scope of this exploratory paper.
Dedicated studies for reducing the fake $\gamma$ background
 as well as those for matching the matrix element to the parton shower may be required,
 because of the peculiar final state kinematics with very large di-jet invariant mass and relatively small
 transverse momenta of jets and a photon, which is required to establish the null radiation zone
 in the $u u \rightarrow \ u u \gamma$ subprocess at the LHC.
It may be possible to take advantage of the technique for identifying the charge of a jet, 
 which has been successfully applied by ALEPH, DELPHI and OPAL 
 when measuring the forward-backward asymmetry of the up quark and that of the down (and strange) quark 
 at the $Z$ boson pole \cite{ado}.  
Although the usefulness of the jet charge measurement
 for much more energetic jets at hadron colliders should yet to be demonstrated, 
 if they can be used to obtain a sample of events
 which are enriched by di-jets with the same sign charges, 
 the signal can be further enhanced 
 because the cancellation due to constructive interference 
 in events with the opposite charge jets will be reduced 
 and also because the major background from the quark-gluon collision events will be reduced.
\\

\end{document}